\documentclass[twocolumn,preprintnumbers,superscriptaddress,prl]{revtex4}
\usepackage{graphicx}
\usepackage{times}
\usepackage{epsfig}
\usepackage{amsmath}
\usepackage{amsfonts}
\usepackage{amssymb}
\usepackage{url}
\usepackage{subfigure}
\newcommand{\be}{\begin{equation}}
\newcommand{\ee}{\end{equation}}
\newcommand{\bea}{\begin{eqnarray}}
\newcommand{\eea}{\end{eqnarray}}

\begin{document}
\pagestyle{plain}
\title{
WIMP Dark Matter Inflation with Observable Gravity Waves
}
\author{Nobuchika Okada}
\affiliation{
Department of Physics and Astronomy,
University of Alabama,
Tuscaloosa, AL 35487, USA
}
\author{Qaisar Shafi}
\affiliation{
Bartol Research Institute,
Department of Physics and Astronomy,
University of Delaware, Newark, DE 19716, USA
}
\begin{abstract}

We present a successful inflation model based on
 $\lambda \phi^4$ potential in which a Standard Model (SM)
 singlet inflaton $\phi$, with mass of around a TeV or less,
 also plays the role of a weakly interacting scalar dark
 matter particle (WIMP).
The WIMP relic abundance generated after inflation
 is in accord with the current observations.
The spectral index $n_s$ lies within the WMAP 1-$\sigma$ bounds,
 while the Planck satellite may observe the tensor-to-scalar ratio,
 a canonical measure of gravity waves, which
 we estimate lies between 0.003 and 0.007.
An unbroken $Z_2$ parity ensures that the scalar WIMP
 is absolutely stable.

\end{abstract}
\maketitle

The idea that the inflaton, a particle responsible
 for primordial inflation, also may play the role
 of scalar WIMP dark matter is most intriguing \cite{KofmanEtAl}
 and therefore worth pursuing.
Ref.~\cite{LiddleEtAl} attempted to implement this idea
 in chaotic inflation with $m^2 \phi^2$ potential.
However, a satisfactory scenario could not be realized
 which is, to a large extent, related to the fact
 that $m \simeq 10^{13}$ GeV, as demanded by inflation,
 far exceeds the canonical WIMP mass of a TeV or so.
Simply replacing the quadratic potential with a quartic one
 does not help solve the conundrum for in this case
 the scalar spectral index $n_s$ and tensor-to-scalar ratio $r$
 lie outside the WMAP 2-$\sigma$ bounds \cite{WMAP}.

In a recent paper, hereafter called \cite{AxionInflation},
 it was shown that $\lambda \phi^4$ inflation,
 if supplemented by the non-minimal gravitational coupling
 $\xi {\cal R} \phi^2$ between the SM gauge singlet scalar field
 $\phi$ and the curvature scalar ${\cal R}$, yields values of $n_s$
 (scalar spectral index) and $r$ (tensor-to-scalar ratio)
 that are compatible with the WMAP 1-$\sigma$ bounds \cite{WMAP}.
This is to be contrasted with minimal $\lambda \phi^4$ inflation
 whose predictions for $n_s$ and $r$ lie outside
 the WMAP 2-$\sigma$ bounds.
A series of earlier papers
 \cite{Bezrukov:2008dt}-\cite{Einhorn:2009bh}
 have previously raised the possibility that
 the inflaton field $\phi$ could be identified
 with the SM Higgs doublet $H$,
 provided the non-minimal coupling
 $\xi = {\cal O}(10^3)-{\cal O}(10^4)$.
While intriguing, doubt about the viability of
 this identification have been raised 
 in \cite{Naturalness1} \cite{Naturalness2}. 
It stems from the observation that for $\xi \gg 1$,
 the energy scale $\lambda^{1/4} m_P /\sqrt{\xi}$
 of inflation exceeds the effective ultraviolet cutoff scale
 $m_P/\xi$, with $m_P$ being the reduced Planck scale,
 assuming the SM Higgs quartic coupling $\lambda$ is of order unity.
In \cite{AxionInflation} we easily evade this problem by making
 $\phi$ a SM gauge singlet field so that the parameter $\lambda$
 is not all that strongly constrained.
Indeed, one finds that consistent with the WMAP 1-$\sigma$ bounds
 on $n_s$ and $r$, $\lambda$ and $\xi$ can lie
 within the relatively wide range,
 $10^{-12} \lesssim \lambda \lesssim 10^{-4}$, and
 $10^{-3} \lesssim \xi \lesssim 10^2$.

In a separate development, it has been noted by several authors \cite{SDM}
 that a stable SM singlet scalar particle, with mass
 $\sim m_h/2 - 1$ TeV, is a viable cold dark matter candidate (WIMP),
 provided it has suitable interactions with the SM Higgs doublet $H$
 and possibly additional fields.
The interaction term $g^2 \phi^2 |H|^2$ plays an especially
 important role in these considerations.
Recent estimates suggest \cite{KMNO} that
 with $g^2 \simeq 0.1$ and dark matter mass $\sim 1$ TeV,
 the relic WIMP abundance is compatible with the value
 $\Omega_{\rm CDM} h^2 = 0.1131 \pm 0.0034$
 determined by WMAP \cite{WMAP}.
This parameter region will be further explored
 in the ongoing and planned direct detection
 experiments of dark matter particle.

In this letter we propose a successful and relatively simple
scenario
 of WIMP dark matter inflation by merging together ideas 
 from \cite{AxionInflation} and \cite{KMNO}. 
Following \cite{AxionInflation}, we employ non-minimal quartic inflation
 in which a gravitational coupling of the inflaton to the curvature
 scalar is included.
The model has a further restriction arising
 from the relic dark matter abundance.
It is shown in \cite{KMNO} that for TeV mass WIMP dark matter,
  the coupling strength $g^2$ must be of order 0.1 or so.
In our case this means that due to radiative corrections involving $g^2$,
 the 'effective' quartic coupling is of order $10^{-3}$.
An important consequence of this WIMP driven inflation
 model is that it predicts both $n_s$ and $r$ in a fairly narrow range.
In particular, $r$ values close to 0.007 may be accessible 
 to Planck satellite searches.
Another important feature of our model is the appearance of
 thermal dark matter relic abundance which arises during preheating
 and subsequent transition to a radiation dominated universe
 with temperature close to $10^7$ GeV.
The energy in the oscillating inflaton field is by then
 essentially negligible.

Consider the following tree level action in the Jordan frame:
\bea
\label{action1}
&& S_J^{tree} = \int d^4 x \sqrt{-g}
 \left[- \left( \frac{m_P^2 + \xi \phi^2}{2} \right) \cal{R}
  \right.  \nonumber \\
&+& \frac{1}{2} (\partial_\mu \phi)^2
  - \left(
    \frac{m_\phi^2}{2} \phi^2 + \frac{\lambda}{4} \phi^4
    \right)
  - \left. \frac{g^2}{2} \phi^2 |H|^2 \right],
\eea
 where $m_P=2.4 \times 10^{18}$ GeV is the reduced Planck mass.
Here we have introduced a $Z_2$ parity under which $\phi$ is odd,
 while the SM fields are all even.
Hence the scalar $\phi$ is stable and will play
 the role of both inflaton and dark matter particle.

First we consider the non-minimal $\lambda \phi^4$ inflation
 in this model \cite{AxionInflation}.
During inflation, with field values close to $m_P$, 
 $\lambda \phi^4$ dominates the scalar potential. 
The relevant one-loop renormalization group 
 improved effective potential \cite{RGEIP} is 
$ V_{\rm eff} = \frac{1}{4} \lambda(t) G(t)^4 \phi^4$,
where $t=\ln(\phi/m_\phi)$, and $G(t)= \exp(- \, \int_0^t dt'
 \gamma(t')/(1+\gamma(t')))$, with $\gamma(t)$
 being the anomalous dimension of the inflaton field.
We employ a leading-log approximation for the effective potential 
\bea
\label{Veff}
  V_{\rm eff}(\phi) \simeq
  \frac{1}{4}
  \left( \lambda_0 + \frac{g^4}{8 \pi^2}
    \ln \left[ \frac{\phi}{m_\phi}\right]
  \right) \phi^4,
\eea
where $\lambda_0=\lambda(t=0)$, $\lambda_0 \ll g^2$, 
 and we have taken $m_\phi$ as the renormalization scale.
In the Einstein frame with a canonical gravity sector,
 the kinetic energy of $\phi$ can be made canonical 
 by defining a new field $\sigma$ \cite{SHW}, 
\bea
\left(\frac{d \sigma}{d \phi}\right)^{-2} =
 \frac{\left( 1 + \frac{\xi \phi^2}{m_P^2} \right)^2}
 {1+(6 \xi +1)\frac{\xi \phi^2}{m_P^2}}.
 \label{kinetic}
\eea
The effective potential in the Einstein frame is then given by
\bea
 V_E(\phi) = \frac{V_{\rm eff}(\phi)}
  {\left(1+\frac{\xi\,\phi^2}{m_P^2}\right)^2}.
\label{potrgi}
\eea

The inflationary slow-roll parameters are given by
\bea
\epsilon(\phi)&=&\frac12 m_P^2 \left({V_E'\over V_E \sigma'}\right)^2,
 \nonumber \\
\eta(\phi)&=&m_P^2\left[
{V_E''\over V_E (\sigma')^2}
\!-{V_E'\sigma''\over V_E (\sigma')^3}\right], \nonumber \\
\zeta^2 (\phi) &=& m_P^4 \left({V_E'\over V_E \sigma'}\right)
\left( \frac{V_E'''}{V_E (\sigma')^3}
 -3 \frac{V_E'' \sigma''}{V_E (\sigma')^4} \right. \nonumber \\
&+& \left. 3 \frac{V_E' (\sigma'')^2}{V_E (\sigma')^5}
- \frac{V_E' \sigma'''}{V_E (\sigma')^4} \right) ,
\eea
where a prime denotes a derivative with respect to $\phi$.
The slow-roll approximation is valid as long as the conditions
 $\epsilon \ll 1$, $|\eta| \ll 1$ and $\zeta^2 \ll 1$ hold.
In this case the scalar spectral index $n_{s}$,
 the tensor-to-scalar ratio $r$, and the running of the spectral index
 $\alpha=\frac{d n_{s}}{d \ln k}$ are approximately given by
\bea
 n_s &\simeq& 1-6\,\epsilon + 2\,\eta, \nonumber \\
 r &\simeq& 16\,\epsilon,  \nonumber \\
 \alpha &=& \frac{d n_{s}}{d \ln k} \simeq
 16\,\epsilon\,\eta - 24\,\epsilon^2 - 2\,\zeta^2.
\label{reqn}
\eea
The number of e-folds after the comoving scale $l$ has
 crossed the horizon is given by
\bea
N_l={1\over \sqrt{2}\, m_P}\int_{\phi_{\rm e}}^{\phi_l}
{d\phi\over\sqrt{\epsilon(\phi)}}\left(\frac{d\sigma}{d \phi}\right)\,,
\label{Ne}
\eea
where $\phi_l$ is the field value at the comoving scale
$l$, and $\phi_e$ denotes the value of $\phi$ at
the end of inflation, defined by
max$(\epsilon(\phi_e) , |\eta(\phi_e)|,\zeta^2(\phi_e)) = 1$.
The amplitude of the curvature perturbation
 $\Delta_{\mathcal{R}}$ is given by
\bea
 \label{Delta}
 \Delta_{\mathcal{R}}^2 = \left. \frac{V_E}
 {24\,\pi^2 \, m_P^2\,\epsilon } \right|_{k_0},
\eea
 which should satisfy the WMAP normalization,
 $\Delta_{\mathcal{R}}^2 = (2.43\pm 0.11)\times 10^{-9}$ \cite{WMAP},
 at $k_0 = 0.002\, \rm{Mpc}^{-1}$.

\begin{figure}[htbp]
\begin{center}
{\includegraphics[width=0.9\columnwidth]{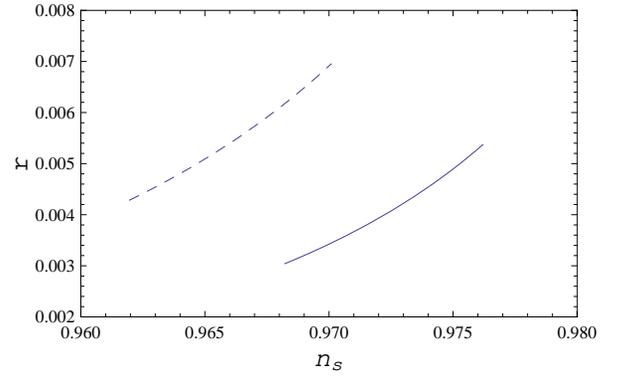}}
\caption{
$r$ vs. $n_s$
 with $N_0 = 60$ (solid curve) and $N_0 = 50$ (dashed curve) e-foldings.
Both curves lie within the WMAP 1-$\sigma$
 (68\% confidence level) bounds.
}
\label{FigNsVSr}
\end{center}
\end{figure}

Using Eqs.~(\ref{Veff})-(\ref{Delta})
 we can obtain various predictions of the radiatively
 corrected non-minimal $\lambda \phi^4$ inflation model.
Once we fix the parameters $\xi$ and the number of e-foldings $N_0$,
 we can predict $n_s$, $r$, and $\alpha = \frac{d n_{s}}{d \ln k}$.
Note that with $m_\phi \simeq 1$ TeV, the coupling $g^2 \simeq 0.1$ 
 in order for the relic density of dark matter 
 to be compatible with the WMAP observations \cite{KMNO}.
In our analysis, we set $g^2=0.1$ and $m_\phi=1$ TeV
 as reference values.
If $\lambda_0 \lesssim g^4/(8 \pi^2)$,
 the potential during inflation is dominated
 by the radiatively corrected part.
We impose  $\lambda_0 \geq 0$  for an unbroken $Z_2$ parity.

\begin{figure}[htbp]
\begin{center}
{\includegraphics[width=0.9\columnwidth]{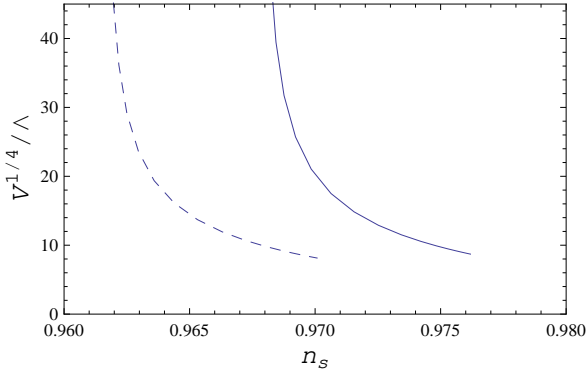}
\caption{
$V^{1/4}/\Lambda$ vs. $n_s$
 with $N_0 = 60$ (solid curve)
 and $N_0 = 50$ (dashed curve) e-foldings.
}
\label{FigNaturalness}
}
\end{center}
\end{figure}

The predicted values of $n_s$ and $r$ are shown
 in Figure~\ref{FigNsVSr} for the number of e-foldings
 $N_0 = 60$ (solid curve) and $N_0 = 50$ (dashed curve).
In non-minimal $\lambda \phi^4$ inflation \cite{AxionInflation},
 the (effective) scalar quartic coupling becomes larger
 according to $\xi$ values, and
 $n_s$ and $r$ approach their asymptotic values,
 $n_s \simeq 0.968$ and $ r \simeq 0.0030$  for $N_0 = 60$
 and $n_s \simeq 0.962$ and $ r \simeq 0.0042$
 for $N_0 = 50$.
These values correspond to the left edge of
 each curve in Figure~\ref{FigNsVSr}.
In the present case, for $\lambda_0 \lesssim g^4/(8 \pi^2)$,
 the radiatively induced term in the effective potential
 dominates the scalar potential and thus
 the effective quartic coupling has a minimum value.
In the limit $\lambda_0 =0$, $n_s$ and $r$ approach
  $n_s \simeq 0.976$ and $ r \simeq 0.0054$
  for $N_0 = 60$
 ($n_s \simeq 0.970$ and $ r \simeq 0.0069$ for $N_0=50$),
 which correspond to the right edge of each curve 
 in Figure~\ref{FigNsVSr}.
We find that the running of the spectral index
 $\alpha=\frac{d n_{s}}{d \ln k}$ very weakly depends on $n_s$, 
 and $\alpha \simeq -0.0005$ ($-0.00075$) 
 for $N_0=60$ ($N_0=50$). 
In Figure~\ref{FigNaturalness}, we show the ratio 
 of the inflation energy scale
 ($V^{1/4}$) to the effective ultraviolet cutoff scale
 ($\Lambda = m_P/\xi$).
This ratio becomes larger as $\xi$ is raised.
The minimum value for this ratio, $V^{1/4}/\Lambda \simeq 8.7$,
 is achieved for $\lambda_0 \lesssim g^4/(8\pi^2)$
 with $\xi \simeq 2000$.
This value marginally exceeds the proposed naturalness bound
 $V^{1/4}/\Lambda < {\cal O}(1)$ \cite{Naturalness2}.
As a more conservative bound, we examine the constraint  
 on the ratio of the Hubble parameter and the effective cutoff 
 scale from the validity of the classical inflationary treatments 
 \cite{Naturalness1}, namely, $ \sqrt{\lambda} \ll H/\Lambda \ll 1$. 
Since $ H/\Lambda \simeq (V^{1/4}/\Lambda)^2 (\Lambda/m_P)$, 
 we find the ratio $\simeq (8.7)^2/2000 \simeq 0.04$ 
 for $\xi \simeq 2000$, 
 while $\sqrt{\lambda} \simeq \sqrt{g^4/(8\pi^2)} \simeq 0.01$. 
Thus, this bound is satisfied.

After inflation, the inflaton field starts to oscillate
 around its potential minimum.
During this period the energy density of the inflaton is
 transmitted to other relativistic fields.
Since a single inflaton cannot decay into other particles
 because of $Z_2$ parity,
 preheating \cite{KofmanEtAl} via, in our case,
 the coupling $g^2 \phi^2 |H|^2$ plays
 the crucial role in energy transmission.
In the first stage of preheating, the inflaton energy density
 is transmitted to $\phi$ particles by explosive production
 through parametric resonance effects
 with $\lambda \phi^4$ potential.
The amplitude of the inflaton filed is reducing its amplitude by
 $\phi$ particle production and the expansion of the universe.
When this becomes smaller than $m_\phi/\sqrt{\lambda}$,
 the term $m_\phi^2 \phi^2$ in the scalar potential dominates
 and the Higgs doublets are explosively produced
 by the (broad) parametric resonance through the coupling
 $g^2 \phi^2 |H|^2$.
This preheating process ends when the amplitude
 becomes smaller than $m_\phi/g$ \cite{KofmanEtAl}.

Since the inflaton is stable, its oscillations
 continue without further energy transfer
 after the preheating era.
The oscillating mode has the equation of state of dust 
 and so, in principle, it can play the role of dark matter.
The present ratio  of the average energy density of
 this oscillation mode to the number density of photons
 is estimated as \cite{LiddleEtAl}
\bea
 \xi_{{\rm dm},0} \simeq 0.44
 \left( \frac{m_\phi}{m_P}\right)^{1/2}
 \left( \frac{\phi_*}{m_P}\right)^2 m_P,
\eea
 where $\phi_*$ is the oscillation amplitude
 at time $t_*$ when $m_\phi=H_*$, the Hubble parameter.
Since $m_\phi > H$ for $t > t_*$,
 $\phi_*$ is basically also the amplitude of
 the oscillating inflaton today. 
With $\phi_* = m_\phi/g $, we find
 $\xi_{{\rm dm},0} \simeq 1.5 \times 10^{-37} m_P$
 for $m_\phi=1$ TeV and $g^2=0.1$.
Comparing it to the observed value by WMAP \cite{WMAP},
 $ \xi_{{\rm dm},0} \simeq 1.1 \times 10^{-27} m_P$, 
 we conclude that the oscillating mode
 has a negligible contribution to the energy density of
 the present universe.
It is worth recalling that successful $m_\phi^2 \phi^2$
 chaotic inflaton requires $m_\phi \simeq 10^{13}$ GeV, 
 and so $g \simeq 10^7$ is needed to realize the observed value
 of $\xi_{{\rm dm},0}$ \cite{LiddleEtAl}! 
In our case, successful inflation is realized by
 non-minimal $\lambda \phi^4$ inflation
 and the inflaton mass plays no role during inflation
 for $m_\phi \ll 10^{13}$ GeV.

We have seen that the energy density 
 in the remnant inflaton oscillations is 
 tiny compared to the observed dark matter relic density.
However, the $\phi$ particle can be a suitable
 WIMP dark matter candidate if the universe is thermalized
 with the reheating temperature high enough for the $\phi$ particle
 to be in thermal equilibrium.
In the preheating scenario, thermalization of the universe
 takes place through decays and multiple scatterings of
 particles (SM Higgs doublets in our model),
 during explosively produced preheating.
A reasonable estimate for the reheating temperature is \cite{KofmanEtAl}
\bea
  T_R \sim 0.5 \sqrt{\Gamma_h m_P},
\eea
 where $\Gamma_h$ is the total decay width of the Higgs boson.
For a relatively light Higgs boson with mass $m_h=120$ GeV for example,
 the dominant decay mode is $h \to b \bar{b}$, so that
\bea
  \Gamma_h \sim \frac{3}{8 \pi} \left( \frac{m_b}{v}\right)^2 m_h,
\eea
 where $m_b \simeq 3$ GeV is the appropriate bottom quark mass,
 and $v=246$ GeV is the VEV of the SM Higgs doublet.
We find $T_R \sim 10^7$ GeV, which is four orders
 of magnitude larger than $m_\phi$($\simeq 1$ TeV).
Thus, we expect that the temperature of the universe
 is high enough for $\phi$ particles to be in thermal equilibrium,
 and the standard WIMP dark matter scenario
 consistent with the WMAP observations
 can be realized, as shown in recent analysis \cite{KMNO}.

In our analysis above, we set $m_\phi=1$ TeV in order
 to keep the inflaton mass much larger
 than the SM Higgs boson mass, $m_\phi \gg m_h$.
This parameter choice makes the preheating process
 as effective as possible.
In general, the preheating process can be reasonably efficient
 even for $m_\phi \sim m_h$ \cite{KofmanEtAl}.
If the inflaton mass can be lowered close to half of the Higgs boson mass,
 the magnitude of the coupling $g^2$ needed to obtain
 the correct dark matter relic abundance becomes
 significantly smaller than 0.1 \cite{KMNO}.
In this case, the radiative corrections to the potential
 is negligible and we can easily obtain $V^{1/4}/\Lambda < 1$
 for a successful inflation scenario
 as shown in \cite{AxionInflation}.

In summary, we have shown that the inflaton and WIMP dark matter can
 indeed be one and the same particle.
In the simplest model this is achieved by supplementing
 the SM with a stable gauge singlet scalar field.
The model overcomes serious challenges faced 
 by chaotic $m^2 \phi^2$ inflation \cite{LiddleEtAl} 
 and, in addition, turns out be quite predictive.
Its dark matter properties will be seriously examined by the ongoing
 direct detection searches \cite{CDMS} \cite{XENON100}.
As far as inflation is concerned the predictions for $n_s$ and $r$
 lie within the WMAP 1-$\sigma$ bounds. With an upper bound
 of around 0.007 on $r$, the model can be excluded
 if the Planck satellite observes values that are significantly
 larger than this.

Finally, one promising extension of our model is
 to introduce an SU(2) triplet scalar field
 with unit SM hypercharge.
This would nicely incorporate neutrino masses and mixings
 via the type-II seesaw mechanism \cite{typeII}. 
Interestingly, this extension essentially coincides
 with a model proposed in \cite{GOS},
 and it can also account for the anomalous cosmic-ray positron flux
 reported by the PAMELA satellite experiment \cite{PAMELA}.
In addition, as analyzed in detail in \cite{GOS2},
 in the type-II seesaw extension of the SM,
 the vacuum stability bound on the Higgs boson mass
 can be reduced to coincide with the current experimental 
 lower bound of 114.4 GeV \cite{LEP2}.

\begin{center}
{\bf Note Added}
\end{center}
Although our approach in unifying the inflaton and dark matter 
 particle is inspired by Refs.~\cite{KofmanEtAl} \cite{LiddleEtAl}, 
 by the recent analysis of SM singlet scalar dark matter \cite{KMNO}, 
 and by non-minimal $\lambda \phi^4$ inflation \cite{AxionInflation}, 
 the model presented in this letter turns out to be identical 
 to the ones proposed in Refs.~\cite{Lerner:2009xg} 
 and \cite{Park}.
In \cite{Lerner:2009xg}, a comprehensive study of certain 
 aspects of this model were presented, and where 
 there is overlap with our work, the results appear 
 to be in broad agreement. 
However, we emphasize that there are several new and important 
 results in this letter. 
Thus, (1) we have emphasized the WMAP constraints 
 on the coupling $g^2$, arising from the thermal relic density
 of dark matter, and evaluated its direct impact 
 on the effective inflaton quartic coupling. 
As a result, our inflationary scenario predicts 
 both $n_s$ and $r$ in a narrow range. 
(2) This feature also then plays an important role 
 in discussing the naturalness of the inflationary scenario 
 as shown in Figure \ref{FigNaturalness}. 
(3) We have considered the preheating scenario after inflation, 
 following \cite{KofmanEtAl} and \cite{LiddleEtAl}, 
 which allows the transition to a radiation dominated universe.
Furthermore, we have shown that the energy density in the remnant
 inflaton oscillations, which plays the role of dark matter 
 in the original scenario \cite{LiddleEtAl}, 
 can be ignored. 
(4) Considering thermalization of the universe via preheating, 
 we have estimated the reheating temperature and shown that 
 it is high enough for the singlet scalar 
 to be in thermal equilibrium. 
This is crucial for a successful WIMP dark matter scenario.

\begin{center}
{\bf Acknowledgments}
\end{center}
N.O. would like to thank the Particle Theory Group
 of the University of Delaware for hospitality during his visit.
The work of Q.S. is supported in part by the DOE
 under grant No. DE-FG02-91ER40626.

%

\end{document}